\documentclass[twocolumn,showpacs,superscriptaddress,prl]{revtex4}

\usepackage{amsbsy,amssymb,amsmath,bm,graphicx}

\begin{document}

\title{Spontaneous current generation in gated nanostructures}

\author{D.~W.~Horsell}
\affiliation{School of Physics, University of Exeter, Exeter, EX4 4QL, UK}
\author{A.~K.~Savchenko}
\affiliation{School of Physics, University of Exeter, Exeter, EX4 4QL, UK}
\author{Y.~M.~Galperin}
\affiliation{Department of Physics, University of Oslo, PO Box 1048 Blindern,
0316 Oslo, Norway} \affiliation{A. F. Ioffe Physico-Technical Institute of
Russian Academy of Sciences, 194021 St.~Petersburg, Russia}
\affiliation{Argonne National Laboratory, 9700 S. Cass av., Argonne, IL 60439,
USA}
\author{V.~I.~Kozub}
\affiliation{A. F. Ioffe Physico-Technical Institute of Russian Academy of
Sciences, 194021 St.~Petersburg, Russia} \affiliation{Argonne National
Laboratory, 9700 S. Cass av., Argonne, IL 60439, USA}
\author{V.~M.~Vinokur}
\affiliation{Argonne National Laboratory, 9700 S. Cass av., Argonne, IL 60439,
USA}
\author{D.~A.~Ritchie}
\affiliation{Cavendish Laboratory, University of Cambridge, Cambridge, CB3 0HE,
UK}

\begin{abstract}
We have observed an unusual dc current spontaneously generated in the
conducting channel of a short-gated GaAs transistor. The magnitude and
direction of this current critically depend upon the voltage applied to the
gate. We propose that it is initiated by the injection of hot electrons from
the gate that relax via phonon emission. The phonons then excite secondary
electrons from asymmetrically distributed impurities in the channel, which
leads to the observed current.
\end{abstract}


\maketitle

Gated semiconductor nanostructures have become the staple diet of modern
condensed matter research and applications. Their small size has resulted in a
wealth of new phenomena observed in electron transport, including universal
conductance fluctuations \cite{LeePRL55,AltshulerJETPL41} and the
photo-galvanic effect \cite{FalkoJETP68,BykovJETPL49}. In such a structure, at
low temperatures we observe a dc current through the conducting channel in the
absence of any applied bias. This current is dependent upon the gate voltage
$V_g$, which dictates its magnitude and direction through the channel.

It was found that the observed current could not be produced by conventional
sources of residual bias and stray interference coupling to the system
\cite{VeselagoJETPL44,KopevSPS23}. We propose a model that eliminates this
apparent ``Maxwell's demon'' required to support the voltage across the sample.
A small gate leakage current is magnified in the source--drain circuit due to
phonon-assisted excitation of localized electrons. While the leakage current
itself has a smooth dependence on $V_g$, the ``spontaneous'' current changes
its direction due to the $V_g$-dependent asymmetry of the channel. It
transpires that the effect is greatest for channels of length $\sim 0.1
\,\mathrm{\mu m}$, which is the key size in contemporary nanostructures.

The experiments were carried out on a GaAs based transistor. The wafer consists
of a $1\cdot10^{17} \,\mathrm{cm^{-3}}$ silicon doped layer
$1450\,\mathrm{\AA}$ thick on an undoped GaAs substrate. A metallic (Au) gate,
of length $0.15 \,\mathrm{\mu m}$ in the current direction and width $9
\,\mathrm{\mu m}$, was formed between the source and drain, see Fig.~1(inset).
For large negative gate voltages the two-terminal conductance of the device is
dominated by the region under the gate; this region defines the ``channel'',
and regions outside the gate are the ``contacts''. Measurements were carried
out in a dilution refrigerator at a base temperature of $30 \,\mathrm{mK}$,
housed in a screen room to suppress external interference. Both the ac and dc
currents were measured via a battery-powered EG\&G 181 low-noise pre-amplifier
within the room.

\begin{figure}[h]
\centerline{
\includegraphics[width=.9\columnwidth]{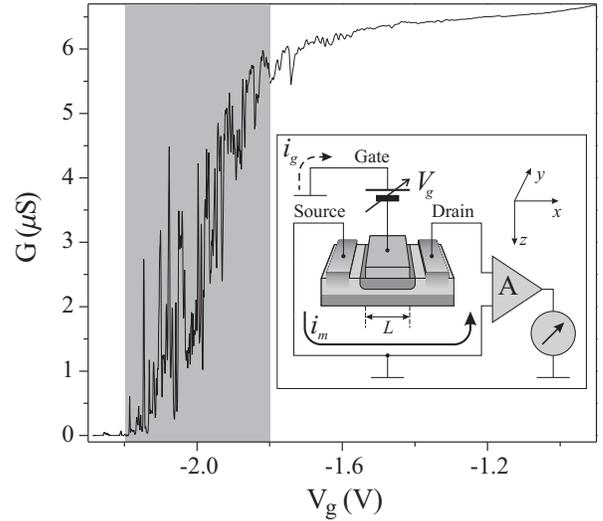}
}
\caption{The two-terminal (differential) conductance as a function of the gate
voltage. The shaded box delimits the range of $V_g$ where the spontaneous
current is resolved. Inset: Circuit (simplified) used to measure the dc current
$i_m$ via the pre-amplifier (A). The   transistor is depicted schematically:
conductive regions are shown in light grey, and depleted regions in dark grey.
\label{fig:1}}
\end{figure}

Figure~1 shows the conductance ($\mathrm{d}i/\mathrm{d}V$) through the channel
as a function of the applied gate voltage. Strong, reproducible structure can
be seen to occur near the pinch-off, associated with mesoscopic hopping and
tunnelling processes \cite{LaikoJETP66}. In the absence of a voltage source in
the source--drain circuit, shown in Fig.~1(inset), one would expect the
measured dc current $i_m$ to be zero at any gate voltage. Contrarily, Fig.~2(a)
shows that a large current occurs that changes direction and magnitude as a
function of $V_g$. This current is only resolved in a small range of gate
voltages, highlighted in Fig.~1, where fluctuations of the conductance are
large.

\begin{figure}[h]
\centerline{
\includegraphics[width=.9\columnwidth]{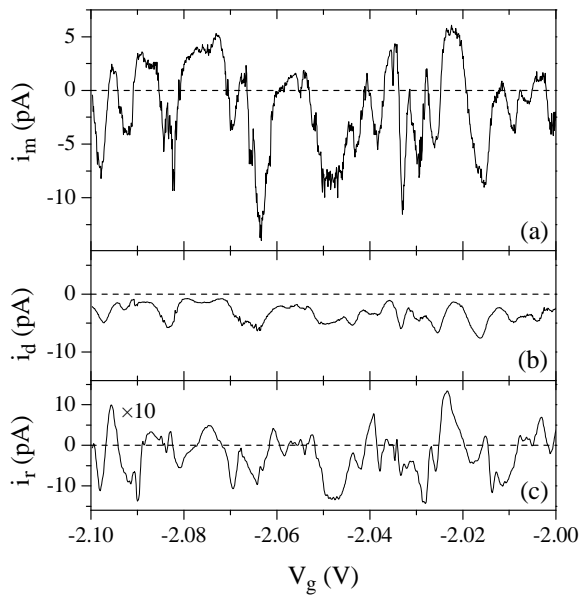}
}
\caption{(a) The measured dc current $i_m$ in the source--drain circuit with no
applied bias within the range of $V_g$ highlighted   in Fig.~1. (b,c) The
contribution to $i_m$ from the drift current $i_d$ due to the presence of
unintentional dc biases in the circuit (b), and from the rectified current
$i_r$ due to the rectification of stray interference (c). \label{fig:2}}
\end{figure}

Three conventional sources of dc current exist in the circuit and can
contribute to $i_m$. First, an unintentional (drift) dc bias $V_d$ can produce
a ``drift'' current $i_d$. Second, rectification of stray interference of
frequency $\omega$ and magnitude $V_\omega$ can produce a rectified current
$i_r$. Third, there is a small leakage current from the gate $i_g$, which
splits into $(1-\alpha)i_g$ in the gate--source and $\alpha i_g$ in the
gate--drain branches of the circuit. If we include an additional, unknown
current $i_0$ as a fourth contribution, the total measured current $i_m$ can be
written as:
\begin{equation}
i_m=\frac{\mathrm{d}i}{\mathrm{d}V}V_d
+\frac{1}{4}\frac{\mathrm{d}^2i}{\mathrm{d}V^2}\sum_{\omega \in
  \Omega} V_{\omega}^2 +\alpha i_g +i_0\;,
\label{eqn:1}
\end{equation}
where the first two terms on the right define $i_d$ and $i_r$ respectively. We
have found that the conventional contributions do not constitute the primary
element of the measured current $i_m$, either in the magnitude or fine
structure, and instead $i_0$ dominates $i_m$. In order to investigate the
nature of $i_0$, we first compare it to $i_d$, $i_r$ and $i_g$; then we propose
a new mechanism of current generation that accounts for its existence.

The first contribution to $i_m$ is derived from an unintentional dc voltage in
the source--drain circuit, which is due to the pre-amplifier. This voltage was
found to change monotonically over the course of an experiment
($1-2~\mathrm{hours}$) by $\sim 200\,\mathrm{nV/hour}$. In Fig.~2(b) the
resultant drift current $i_d$ that can be ascribed to the average of this
voltage, $V_d$, is shown (Eqn.~1). From this we see that the amplitude of $i_d$
is much smaller than the measured current, and more importantly, it is only
driven in one direction. To quantify this difference, we calculate the
correlation coefficient $C(i_d,i_m)=\langle\delta i_d\delta i_m\rangle/
\langle\delta i_d^2\rangle^{1/2}\langle\delta i_m^2\rangle^{1/2}$, where
$\delta i=i-\langle i\rangle$ and $\langle\cdots\rangle$ is an average over the
gate voltage range shown in the figure. We find $C(i_d,i_m)=0.61$, which
increases to $>0.9$ when an intentional bias voltage ($\sim 1 \,\mathrm{mV}$)
is applied such that $i_d$ dominates $i_m$. However, if all the conventional
contributions are first subtracted from $i_m$ (see below) we find that
$C(i_d,i_0)$ is only $-0.28$. This low correlation, together with the small
magnitude and singular direction shows conclusively that the mechanism
associated with a dc bias voltage cannot account for the observed effect.

The second contribution to the measured current is rectification as a result of
the non-linear nature of the system (as evinced in Fig.~1). The non-linearity
of nanostructures has been observed previously,
e.g.~\cite{LaikoJETP66,TaboryskiPRB49,LinkeScience286}. For such a system, the
rectified current $i_r$ is related, through the second derivative
$\mathrm{d}^2i/\mathrm{d}V^2(V_g)$, to the second harmonic response at
frequency $2\omega$ to an applied ac bias at $\omega$. In the absence of an
applied ac bias, a rectified current is still present due to residual stray
interference coupling to the circuit, predominantly the part outside the screen
room. Therefore $i_r$ as a function of $V_g$ can be reconstructed from
measurements of the second harmonic and the integral of the voltage noise
across the channel. In Eqn.~1, the frequency range $\Omega$ for which
$V_\omega\neq0$ was found experimentally to have an upper limit of $20
\,\mathrm{kHz}$. The calculated rectified current is shown in Fig.~2(c), where
it can be seen that it is approximately an order of magnitude smaller than
$i_m$. If we also compare the correlation $C(i_r,i_0)=0.14$ with that obtained
when a strong ac bias ($V_\omega>100 \,\mathrm{\mu V}$) is intentionally
applied to the channel, $C(i_r,i_m)\sim1$, we conclude that $i_0$ cannot be
related to the rectification of stray interference. We confirmed this in
additional experiments where the measurement apparatus outside the screen room
was replaced by analogue meters and batteries (to control $V_g$) mounted
directly upon the refrigerator inside the room. The measurement of $i_0(V_g)$
by discrete points (the meters being read by candlelight) yielded the same
result as that presented.

Figure~3 shows the spontaneous current $i_0\approx i_m-i_d-i_r$. Also shown is
the gate leakage current $i_g$. This is roughly constant ($\sim
2\,\mathrm{pA}$) and its contribution to the measured current, determined by
$\alpha\sim0.5$, is small compared to $i_0$. In addition, it is important to
note that $i_0$ flows around the source--drain circuit, whereas $i_g$ flows
down each branch in the same direction, see Fig.~1(inset). The absence of fine
structure in $i_g$ also shows that it cannot directly account for $i_0$,
although in the model we propose it plays a key role in its generation.

\begin{figure}[h]
\centerline{
\includegraphics[width=.9\columnwidth]{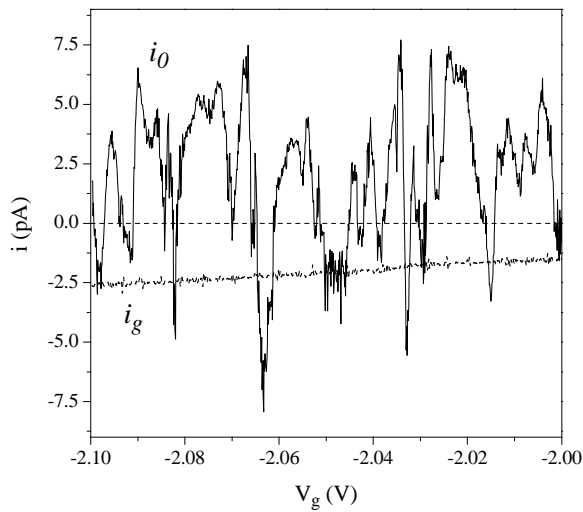}
}
\caption{The spontaneous current $i_0$, after subtracting the contributions of
the drift and rectified currents from the measured current $i_m$ in Fig.~2.
Also shown is the gate leakage current $i_g$ in the same range of gate voltage.
\label{fig:3}}
\end{figure}

It is interesting to note that samples defined by longer gate lengths, up to
$2.0 \,\mathrm{\mu m}$, have shown no evidence in our experiments for producing
spontaneous current. This suggests a critical dependence of its magnitude on
the length of the channel. The model discussed below shows that indeed there
can be an optimal channel length for the observation of the effect, which is
close to that of our experiment.

The model explains the experimental observations in terms of magnification of
the gate leakage current. First we note that, although the leakage current is
small, the power it supplies, $i_gV_g\sim4\cdot10^{-12}\,\mathrm{W}$, is enough
to support the current $i_0$ in the source--drain circuit (in fact, it is
significantly larger than the dissipated power $i_0^2R\sim
10^{-17}\,\mathrm{W}$, where $R$ is the circuit resistance). Thus we do not
have a situation of \textit{perpetuum mobile}.

\begin{figure}[h]
\centerline{
\includegraphics[width=\columnwidth]{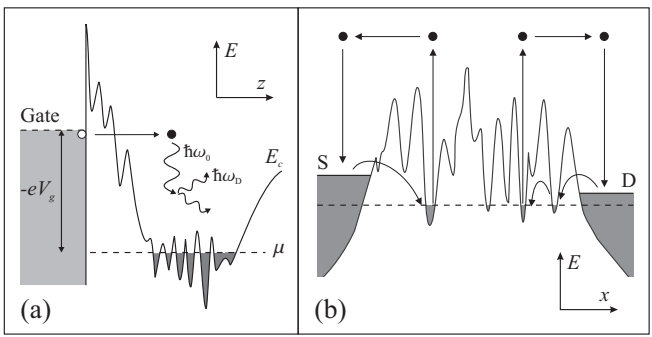}
}
\caption{The physical mechanism to explain the spontaneous current as a
multistage process of relaxation and excitation. (a) The relaxation of hot
electrons from the gate into the channel (Stages a--b, described in the text).
(b) Diffusion and neutralization of secondary electrons from the channel
(Stages c--d). The equilibrium Fermi level is shown as a dotted
line.\label{fig:4}}
\end{figure}

We suggest the following sequence of events (detailed below), shown in
Fig.~\ref{fig:4}: (a) emission of optical phonons by electrons tunnelling from
the gate; (b) conversion of optical into acoustic phonons; (c) excitation of
``secondary'' electrons by these acoustic phonons; (d) diffusion of secondary
electrons into the contacts, and their subsequent return to the channel.

\paragraph{a.}
Electrons from the gate are injected into the channel with energy $\simeq
e|V_g| \approx2 \,\mathrm{eV}$ (Fig.~3), which is large compared with the
sample temperature. So these electrons relax rapidly, predominantly by the
emission of a cascade of $N_p=eV_g/\hbar \omega _0 \gg 1$ \textit{optical}
phonons with energy $\hbar \omega _0$, either inside the channel or in the
contacts close to the channel. The hot electrons mostly reside in the side
valleys of GaAs with small mobility and large effective mass $0.35\cdot
10^{-27} \,\mathrm{g}$ \cite{Smith}, so their initial velocity can be roughly
estimated as $v\sim 10^8\,\mathrm{cm/s}$ (from the condition $mv^2/2\approx 1
\,\mathrm{eV}$). Consequently, in a short channel (in our case $\sim 10^{-5}$
cm) only a few optical phonons are emitted before the hot electron reaches a
contact, where it continues to emit optical phonons. (The contacts are made of
heavily doped GaAs with Fermi energy $\sim 10 \,\mathrm{meV}$, and in such
material the emission of optical phonons remains the most efficient mechanism
of electron energy relaxation~\cite{Gant_Lev}.) The typical size of the contact
region where the phonons are emitted is
$L_{\text{dif}}=(DN_p\tau_{\text{e-ph}})^{1/2}$, where $D=v^2\tau_p/3$ and
$\tau_p$ (estimated below) is the elastic electron mean free time, while
$\tau_{\text{e-ph}}\sim 10^{-13} \,\mathrm{s}$ is  the relaxation time due  to
emission of an optical phonon.

It is known that the room-temperature electron mobility in the side valley is
$\sim 150 \,\mathrm{cm^2V^{-1}s^{-1}}$, from which one estimates for an
electron energy of $k_BT\sim 30 \,\mathrm{meV}$ that $\tau_p\approx 3 \cdot
10^{-14} \,\mathrm{s}$. It is expected that scattering of hot electrons is
mostly due to polar scattering by optical phonons~\cite{Gant_Lev} for which
$\tau_p \propto \varepsilon^{1/2}$. Thus, for a typical energy $\sim 1
\,\mathrm{eV}$ the value of $\tau_p$ is about an order of  magnitude larger
than its equilibrium, room-temperature value. Taking into account an additional
factor $\ln (\varepsilon/\hbar  \omega_0)$ in the relaxation
rate~\cite{Gant_Lev} we estimate $\tau_p\sim 10^{-13}\,\mathrm{s}$.
Consequently, an estimate for  the penetration depth of a hot electron into the
contact is $L_{\text{dif}} \approx 1 \,\mathrm{\mu m}$.

\paragraph{b.}
Each optical phonon quickly decays into two high-energy acoustic phonons over
the characteristic time $\tau_{\text{op}} \sim
10^{-11}\,\mathrm{s}$~\cite{Danilchenko}. However, the decay of acoustic
phonons is much weaker. The transverse modes practically do not decay, and
their relaxation is mostly due to their conversion to longitudinal modes in the
course of phonon--impurity scattering. One expects that the cross-section for
the scattering of transverse acoustic phonons with $\hbar \omega_D \sim 15
\,\mathrm{meV}$ by impurities is of the order of the atomic one,
$\sigma\sim10^{-15} \,\mathrm{cm^2}$. The mean free path for such phonons
within the contacts at impurity concentration $N_i\sim 10^{17}
\,\mathrm{cm^{-3}}$ is $l=(\sigma N_i)^{-1}\sim 10^{-2} \,\mathrm{cm}$, and so
the majority of the phonons created over the distance of $L_{\text{dif}}$ can
easily (ballistically) reach the channel region. Thus, the number of acoustic
phonons in the channel produced by one tunnelling electron is $\sim eV_g/2\hbar
\omega_D$.

\paragraph{c.}
These phonons ionize donors in the channel creating ``secondary'' electrons.
The probability for them to do so can be estimated using Fermi's golden rule,
the squared matrix element being
\begin{equation*}
\frac{\lambda^2 \hbar q}{M\omega_{q}}\left|\left\langle
    \frac{e^{-r/a}}{a^{3/2}}\left|\frac{e^{i{\bf qr}}}{{\cal
          V}^{1/2}}\right| {\frac{e^{i{\bf kr}}}{{\cal
          V}^{1/2}}}\right\rangle\right|^2 \approx
\frac{\lambda^2\hbar a}{M\omega_qq{\cal V}}\,.
\end{equation*}
Here $a$ is the localization length, $\bf q$ and $\bf k$ are the wave vectors
of the phonon and excited electron, respectively; $\lambda$ is the deformation
potential, $M$ the atomic mass, and $\cal V$ is the normalization volume. One
calculates the phonon scattering rate due to the ionization processes as
\begin{equation*}
\frac{1}{\tau_\text{{ph,i}}} \sim \frac{\omega_q\,
    N_ia^3}{(qa)^2}
  \frac{\lambda^2}{E_b^2}\left(\frac{\hbar
  \omega_q-E_i}{E_b}\right)^{1/2}\, ,
\end{equation*}
where $E_i$ is the donor ionization energy and $E_b$ is of the order of the
atomic energy. Since $\lambda \sim E_b$, $qa \sim 10 $, $\hbar \omega_q-E_i
\sim 10^{-2} E_b$ and $N_ia^3 \sim 1$, one has $\tau_{\text{ph,i}}^{-1} \sim
10^{-3}\omega_q$. Correspondingly, the mean free path with respect to
ionization is about $3\cdot10^{-5}\,\mathrm{cm}$.

From the above estimates it follows that the non-equilibrium acoustic phonons
effectively relax within the channel via ionization of the donors. The net
current of secondary electrons is thus $\sim (eV_g/2\hbar\omega_D)i_g$. The
term in parentheses, which is $\sim100$, can be regarded as an amplification
factor for the gate leakage current. Experimentally it was shown above that the
required magnification is $\lesssim10$, which is well within the theoretical
limit.

\paragraph{d.}
Since $\hbar\omega_D \sim 15 \,\mathrm{meV}>E_i$, the secondary electrons have
a large characteristic energy and a correspondingly large velocity, $\sim (2-4)
\cdot 10^{7}$ cm/s, to escape from the initial donor. The energy of these
electrons is well above the conduction band edge, thus they are only weakly
sensitive to the potential landscape of the channel. This fact ensures that
only a small difference exists in the flow of secondary electrons towards the
two contacts. For a characteristic electron energy around $10 \,\mathrm{meV}$
and mean free time $10^{-12}\,\mathrm{s}$ (estimated for the Coulomb scattering
by charged impurities with concentration $N_i$) the mean free path is $\sim
(2-4)\cdot 10^{-5} \,\mathrm{cm}$, which is of the order of the length of the
channel. Hence most of the secondary electrons reach the contacts ballistically
where they relax by electron--electron interaction.

The system now needs to restore quasi-neutrality, and the only way to do so is
for electrons to hop back to the channel and be captured by the ionized donors.
Although the secondary electrons diffuse equally to both contacts, their return
to the donors is \textit{asymmetric}. This is due to the fact that the channel
is mesoscopic, and the hopping paths from the two contacts are different,
Fig.~4(b). As a result, the electrochemical potentials in the contacts are
increased differently with respect to the equilibrium Fermi level. It is this
potential difference that drives the current $i_0$ in the external circuit. In
experiment the degree of asymmetry is controlled by the gate voltage which
determines the spatial position of donors in the channel. Thus the magnitude
and direction of $i_0$ is critically dependent upon its value, Fig.~3.

Our estimations above show that the proposed mechanism is indeed realized in
systems with channel length $0.1 \,\mathrm{\mu m}$; moreover this length
appears to be an optimal one for its realization. Upon increasing the length
the effect is significantly suppressed, both due to the increased probability
for secondary electrons to relax directly back to the ionized donors in the
channel (Stage d), and due to the decrease in the asymmetry of the channel.
Contrarily, in shorter channels the process of ionization (Stage c) will be
less efficient.

In conclusion we have observed a novel ``phonon--electric'' effect in a gated
nanostructure, which is seen as a spontaneous generation of a dc current with
no driving voltage applied. Our explanation is based on the combination of
leakage current magnification mediated by phonons and asymmetry in the channel
controlled by the gate voltage.

AKS is grateful to V.~Pokrovskii, V.~I.~Falko, M.~Entin, M.~E.~Raikh, and
A.~O.~Orlov for useful discussions. This work was supported by the EPSRC (U.K.)
and the U.S. Department of Energy Office of Science through contract No.
W-31-109-ENG-38.


\begin{thebibliography}{12}
\expandafter\ifx\csname natexlab\endcsname\relax\def\natexlab#1{#1}\fi
\expandafter\ifx\csname bibnamefont\endcsname\relax
  \def\bibnamefont#1{#1}\fi
\expandafter\ifx\csname bibfnamefont\endcsname\relax
  \def\bibfnamefont#1{#1}\fi
\expandafter\ifx\csname citenamefont\endcsname\relax
  \def\citenamefont#1{#1}\fi
\expandafter\ifx\csname url\endcsname\relax
  \def\url#1{\texttt{#1}}\fi
\expandafter\ifx\csname urlprefix\endcsname\relax\def\urlprefix{URL }\fi
\providecommand{\bibinfo}[2]{#2}
\providecommand{\eprint}[2][]{\url{#2}}

\bibitem[{\citenamefont{Lee and Stone}(1985)}]{LeePRL55}
\bibinfo{author}{\bibfnamefont{P.~A.} \bibnamefont{Lee}} \bibnamefont{and}
  \bibinfo{author}{\bibfnamefont{A.~D.} \bibnamefont{Stone}},
  \bibinfo{journal}{Phys. Rev. Lett.} \textbf{\bibinfo{volume}{55}},
  \bibinfo{pages}{1622} (\bibinfo{year}{1985}).

\bibitem[{\citenamefont{Altshuler}(1985)}]{AltshulerJETPL41}
\bibinfo{author}{\bibfnamefont{B.~L.} \bibnamefont{Altshuler}},
  \bibinfo{journal}{JETP Lett} \textbf{\bibinfo{volume}{41}},
  \bibinfo{pages}{648} (\bibinfo{year}{1985}).

\bibitem[{\citenamefont{Fal'ko and Khmel'nitskii}(1989)}]{FalkoJETP68}
\bibinfo{author}{\bibfnamefont{V.~I.} \bibnamefont{Fal'ko}} \bibnamefont{and}
  \bibinfo{author}{\bibfnamefont{D.~E.} \bibnamefont{Khmel'nitskii}},
  \bibinfo{journal}{Sov. Phys. JETP} \textbf{\bibinfo{volume}{68}},
  \bibinfo{pages}{186} (\bibinfo{year}{1989}).

\bibitem[{\citenamefont{Bykov et~al.}(1989)\citenamefont{Bykov, Gusev, Kvon,
  Lubyshev, and Migal}}]{BykovJETPL49}
\bibinfo{author}{\bibfnamefont{A.~A.} \bibnamefont{Bykov}},
  \bibinfo{author}{\bibfnamefont{G.~M.} \bibnamefont{Gusev}},
  \bibinfo{author}{\bibfnamefont{Z.~D.} \bibnamefont{Kvon}},
  \bibinfo{author}{\bibfnamefont{D.~I.} \bibnamefont{Lubyshev}},
  \bibnamefont{and} \bibinfo{author}{\bibfnamefont{V.~P.} \bibnamefont{Migal}},
  \bibinfo{journal}{JETP Lett.} \textbf{\bibinfo{volume}{49}},
  \bibinfo{pages}{13} (\bibinfo{year}{1989}).

\bibitem[{\citenamefont{Veselago et~al.}(1986)\citenamefont{Veselago,
  Zavaritskii, Nunuparov, and Berkut}}]{VeselagoJETPL44}
\bibinfo{author}{\bibfnamefont{V.~G.} \bibnamefont{Veselago}},
  \bibinfo{author}{\bibfnamefont{V.~N.} \bibnamefont{Zavaritskii}},
  \bibinfo{author}{\bibfnamefont{M.~S.} \bibnamefont{Nunuparov}},
  \bibnamefont{and} \bibinfo{author}{\bibfnamefont{A.~B.}
  \bibnamefont{Berkut}}, \bibinfo{journal}{JETP Lett.}
  \textbf{\bibinfo{volume}{44}}, \bibinfo{pages}{490} (\bibinfo{year}{1986}).

\bibitem[{\citenamefont{Kop'ev et~al.}(1989)\citenamefont{Kop'ev, Nadtochii,
  and Ustinov}}]{KopevSPS23}
\bibinfo{author}{\bibfnamefont{P.~S.} \bibnamefont{Kop'ev}},
  \bibinfo{author}{\bibfnamefont{M.~Y.} \bibnamefont{Nadtochii}},
  \bibnamefont{and} \bibinfo{author}{\bibfnamefont{V.~M.}
  \bibnamefont{Ustinov}}, \bibinfo{journal}{Sov. Phys. Semicond.}
  \textbf{\bibinfo{volume}{23}}, \bibinfo{pages}{694} (\bibinfo{year}{1989}).

\bibitem[{\citenamefont{Laiko et~al.}(1987)\citenamefont{Laiko, Orlov,
  Savchenko, Ilichev, and Poltoratskii}}]{LaikoJETP66}
\bibinfo{author}{\bibfnamefont{E.~I.} \bibnamefont{Laiko}},
  \bibinfo{author}{\bibfnamefont{A.~O.} \bibnamefont{Orlov}},
  \bibinfo{author}{\bibfnamefont{A.~K.} \bibnamefont{Savchenko}},
  \bibinfo{author}{\bibfnamefont{E.~A.} \bibnamefont{Ilichev}},
  \bibnamefont{and} \bibinfo{author}{\bibfnamefont{E.~A.}
  \bibnamefont{Poltoratskii}}, \bibinfo{journal}{Sov. Phys. JETP}
  \textbf{\bibinfo{volume}{66}}, \bibinfo{pages}{1258} (\bibinfo{year}{1987}).

\bibitem[{\citenamefont{Taboryski et~al.}(1994)\citenamefont{Taboryski, Geim,
  Persson, and Lindelof}}]{TaboryskiPRB49}
\bibinfo{author}{\bibfnamefont{R.}~\bibnamefont{Taboryski}},
  \bibinfo{author}{\bibfnamefont{A.~K.} \bibnamefont{Geim}},
  \bibinfo{author}{\bibfnamefont{M.}~\bibnamefont{Persson}}, \bibnamefont{and}
  \bibinfo{author}{\bibfnamefont{P.~E.} \bibnamefont{Lindelof}},
  \bibinfo{journal}{Phys. Rev. B} \textbf{\bibinfo{volume}{49}},
  \bibinfo{pages}{7813} (\bibinfo{year}{1994}).

\bibitem[{\citenamefont{Linke et~al.}(1999)\citenamefont{Linke, Humphrey,
  Lofgren, Sushkov, Newbury, Taylor, and Omling}}]{LinkeScience286}
\bibinfo{author}{\bibfnamefont{H.}~\bibnamefont{Linke}},
  \bibinfo{author}{\bibfnamefont{T.~E.} \bibnamefont{Humphrey}},
  \bibinfo{author}{\bibfnamefont{A.}~\bibnamefont{Lofgren}},
  \bibinfo{author}{\bibfnamefont{A.~O.} \bibnamefont{Sushkov}},
  \bibinfo{author}{\bibfnamefont{R.}~\bibnamefont{Newbury}},
  \bibinfo{author}{\bibfnamefont{R.~P.} \bibnamefont{Taylor}},
  \bibnamefont{and} \bibinfo{author}{\bibfnamefont{P.}~\bibnamefont{Omling}},
  \bibinfo{journal}{Science} \textbf{\bibinfo{volume}{286}},
  \bibinfo{pages}{2314} (\bibinfo{year}{1999}).

\bibitem[{\citenamefont{Smith}(1978)}]{Smith}
\bibinfo{author}{\bibfnamefont{R.~A.} \bibnamefont{Smith}},
  \emph{\bibinfo{title}{Semiconductors}} (\bibinfo{publisher}{Cambridge
  University Press}, \bibinfo{address}{Cambridge, England},
  \bibinfo{year}{1978}), \bibinfo{edition}{2nd} ed.

\bibitem[{\citenamefont{Gantmakher and Levinson}(1987)}]{Gant_Lev}
\bibinfo{author}{\bibfnamefont{V.~F.} \bibnamefont{Gantmakher}}
  \bibnamefont{and} \bibinfo{author}{\bibfnamefont{Y.~B.}
  \bibnamefont{Levinson}}, \emph{\bibinfo{title}{Carrier Scattering in Metals
  and Semiconductors}} (\bibinfo{publisher}{North-Holland},
  \bibinfo{year}{1987}).

\bibitem[{\citenamefont{Danilchenko et~al.}(1994)\citenamefont{Danilchenko,
  Klimashov, Roshko, and Asche}}]{Danilchenko}
\bibinfo{author}{\bibfnamefont{B.}~\bibnamefont{Danilchenko}},
  \bibinfo{author}{\bibfnamefont{A.}~\bibnamefont{Klimashov}},
  \bibinfo{author}{\bibfnamefont{S.}~\bibnamefont{Roshko}}, \bibnamefont{and}
  \bibinfo{author}{\bibfnamefont{M.}~\bibnamefont{Asche}},
  \bibinfo{journal}{Phys. Rev. B} \textbf{\bibinfo{volume}{50}},
  \bibinfo{pages}{5725} (\bibinfo{year}{1994}).
\end{thebibliography}
\end{document}